# Is ZrB$_{12}$ two gap superconductor?

V.A. Gasparov, N.S. Sidorov, I.I. Zver'kova

*Institute of Solid State Physics RAS, Chernogolovka, 142432, Russian Federation*

We report the measurements of the temperature dependence of the resistivity, ρ(T), magnetic penetration depth, λ(T) the lower, H$_{c1}$(T), and upper, H$_{c2}$(T), critical magnetic fields, for single crystals of dodecaboride ZrB$_{12}$, diboride ZrB$_2$ and thin films of diboride MgB$_2$. We observe a number of deviations from conventional behavior in these materials. Although ZrB$_{12}$ behaves like a simple metal in the normal state, the resistive Debye temperature, 300 K, is three times smaller relative to that (800-1200 K) calculated from the specific heat, C(T), data. We observe predominantly quadratic temperature behavior of resistivity in ZrB$_{12}$ below 25 K, and in ZrB$_2$ below 100 K, indicating the possible importance of the electron-electron interaction in these borides. Superfluid density of ZrB$_{12}$ displays unconventional temperature dependence with pronounced shoulder at T/T$_c$ equal to 0.65. Contrary to conventional theories we found a linear temperature dependence of H$_{c2}$(T) for ZrB$_{12}$ from $T_c$ down to 0.35 K. We suggest that both λ(T) and H$_{c2}$(T) dependencies in ZrB$_{12}$ can be explained by two band BCS model with different superconducting gap and $T_c$.

## 1. INTRODUCTION

Recent discovery of superconductivity in magnesium diboride [1] has initiated a substantial interest to potential "high temperature" superconducting transition in other borides [2]. Yet, only nonstoichiometric boride compounds (MoB$_{2.5}$, NbB$_{2.5}$, Mo$_2$B, W$_2$B, BeB$_{2.75}$) demonstrate such transition [3-6]. Absence of superconducting transition in stoichiometric borides is clearly not in line with the old idea about superconductivity in metallic hydrogen [7] recently applied by Kortus *et al.* [8] to explain superconductivity in MgB$_2$. Potential clue to this contradiction my lay in the crystal structure of these boron compounds, in particular in their cluster structure. Crystal structure clearly plays an important role in superconductivity. Although it is widely accepted that the layered structure is crucial for high-$T_c$ superconductivity, one can argue that clusters of light atoms are important for high $T_c$ as well. In particular, there are a number of rather high-$T_c$ superconductors among 3D cluster compounds. Those are alkali metal doped C$_{60}$ compounds (fullerides) Me$_3$C$_{60}$ (Me=K, Na, Rb, Cs) with highest $T_c$ up to 33 K for RbCs$_2$C$_{60}$ [9,10]. It is also known that boron atoms form clusters. These are octahedral B$_6$ clusters in MeB$_6$, icosahedral B$_{12}$ clusters in β- rhombohedral boron, and cubo-octahedral B$_{12}$ clusters in MeB$_{12}$.

The quest for superconductivity in these cluster compounds has a long history. Several superconducting cubic hexaborides - MeB$_6$ and dodecaborides - MeB$_{12}$ have been discovered by Matthias *et al.* back in late 60's [11] (Me=Sc, Y, Zr, La, Lu, Th). Many other cluster borides (Me=Ce, Pr, Nd, Eu, Gd, Tb, Dy, Ho, Er, Tm) were found to be ferromagnetic or antiferromagnetic [11,12]. It was suggested [3] that the superconductivity in YB$_6$ and ZrB$_{12}$ ($T_c$ of 6.5-7.1 K and 6.03 K, respectively [3]) is exactly due to the effect of a cluster of light boron atoms. At the same time, a much smaller isotope effect on $T_c$ for boron in comparison with Zr isotopic substitution suggests that the boron in ZrB$_{12}$ serves as inert background for the Zr-driven superconductivity [13,14]. Clearly systematic study of ZrB$_{12}$ is needed to address the question of superconductivity in this compound.

Superconductivity in ZrB$_{12}$-based compounds was discovered a while ago [11], however there has been little and controversial effort devoted to study of basic superconducting and the electron transport properties of these compounds. In our recent study of the electron transport and superconducting properties of polycrystalline ZrB$_{12}$ [15-18], we demonstrated that this compound behaves like a normal metal with the usual Bloch–Grüneisen dependence of ρ(*T*) but with rather low resistive Debye temperature (T$_R$ =280 K). The latter is almost three times smaller than Debye temperature obtained from C(T) data [19]. We observed linear temperature dependence of λ(T) on polycrystalline samples below T$_c$/2 which could be an evidence of d-wave pairing in this compound. Furthermore, contrary to conventional theories, we found a linear temperature dependence of H$_{c2}$(T). Recently the band structure calculations of ZrB$_{12}$ [20] have been also reported. It was concluded, that the band structure of ZrB$_{12}$ is composed of one open and two closed Fermi surface sheets.

Our data contradict the report of Daghero *et al.* [21] dealing with the point-contact spectroscopy (PCS) of single crystals of ZrB$_{12}$ at temperatures close to $T_c$. In Daghero's report it was concluded that ZrB$_{12}$ is a strong coupling s-wave superconductor, with 2Δ(0)/k$_B$T$_c$=4.7. Tsindlekht *et al.* [22] came to a similar conclusion from tunneling and magnetic char-acterization of $ZrB_{12}$ single crystals at the temperatures also very close to $T_c$ (4.5 K - 6 K). Lortz et al. [19] and Wang et al. [23], reported C(T), $\rho(T)$, magnetic susceptibility, and thermal expansion measurements of $ZrB_{12}$ samples prepared by one of us (VAG) and concluded that it is a single gap marginal BCS superconductor which undergoes transition from type-I superconductor near $T_c$ to type II superconductor below 4.6 K with $2\Delta(0)/k_BT_c=3.7$, the value that is lower than that obtained from PCS [21] and STM data [22]. Large difference in $H_{c2}$ characteristics reported in the above-mentioned papers was discussed in terms of surface superconductivity. Most of the features of $MgB_2$ discovered so far can be explained by two band superconductivity model [24]. We believe that test of the predictions of this model for $ZrB_{12}$ too, may explain observed controversy of published data. Knowledge of the electron transport and superconducting properties in this cluster compound is critical for understanding these conflicting results.

This has been the motivation for current systematic study of the temperature dependencies of $\rho(T)$, $\lambda(T)$, lower $H_{c1}(T)$, and upper $H_{c2}(T)$, critical magnetic fields in single crystals of $ZrB_{12}$. In this report, we confirm unusual superconducting properties of $ZrB_{12}$ and argue that the published results can be reconciled by two-band superconductivity. Comparative data from $ZrB_2$ single crystals and thin films of $MgB_2$ are also presented.

The structure of this paper is as follows. In Section II we report on the samples details and experimental techniques. Section III describes the electron transport in these compounds and Section IV describes the temperature dependence of $\lambda(T)$ in $ZrB_{12}$ samples and $MgB_2$ thin films. The data on $H_{c1}(T)$ and $H_{c2}(T)$ are presented in Section V. Section VI contains our conclusions.

## II. EXPERIMENTAL SETUP

Under ambient conditions, dodecaboride $ZrB_{12}$ crystallizes in the *fcc* structure of the $UB_{12}$ type (space group *Fm3m*, $a=0.74075$ nm [25,26], see Fig.1). In this structure, the Zr atoms are located at interstitial openings among the close-packed $B_{12}$ clusters. In contrast, $ZrB_2$ shows a phase consisting of two-dimensional graphite-like monolayers of boron atoms with a honeycomb lattice structure and the lattice parameters $a=0.30815$ nm and $c=3.5191$ nm (space group *P6/mmm*), intercalated with Zr monolayers [2].

Our $ZrB_{12}$ and $ZrB_2$ single crystals were grown using floating-zone method [17, 18, 25] similar to [26]. The obtained single crystal ingots had a typical diameter of about 5-6 mm and a length of 40 mm. Measured specific weight of the $ZrB_{12}$ rod was 3.60 g/cm$^3$, in a good agreement with the theoretical density. The cell parameter of $ZrB_{12}$, $a=0.74072 \pm 0.00005$ nm, is very close to the published values [26]. To assure good quality of our samples we performed metallographic and X-ray investigations of as grown ingots. We discovered that most parts of the $ZrB_{12}$ ingot contained needle like phase of non-superconducting $ZrB_2$ (see Fig.2). We believe that $ZrB_2$ needles are due to preparation of $ZrB_{12}$ single crystals from a mixture of a certain amount of $ZrB_2$ and an excess of boron [18,25]. Therefore, special care has been taken to cut the samples from $ZrB_2$ phase free parts.

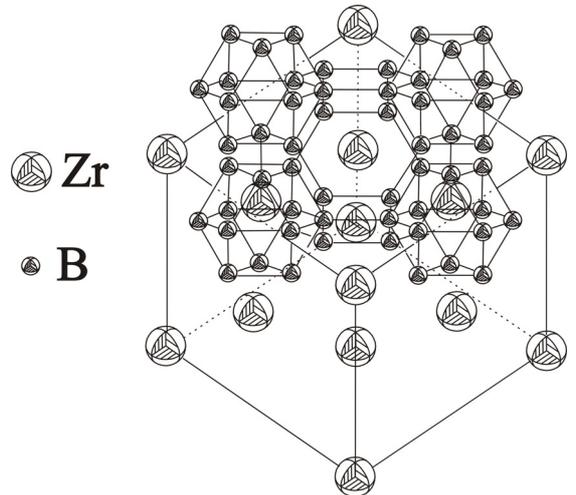

FIG.1. Lattice structure of dodecaboride $ZrB_{12}$. For clarity $B_{12}$ clusters are shown only on the upper face of the lattice.

For this study, two highly crystalline, superconducting films of $MgB_2$ were grown on an *r*-plane sapphire substrate in a two-step process [27]. Deposition of boron precursor films via electron-beam evaporation was followed by *ex-situ* post annealing at 890°C in the presence of bulk $MgB_2$ and Mg vapor. We investigated films of 500 and 700 nm thick with corresponding $T_{c0}$'s of 38 K and 39.2 K, respectively. The details of the preparation technique are described elsewhere [27].

We used spark erosion method to cut the single crystal ingots into rectangular <100> oriented bars of about $0.5 \times 0.5 \times 8$ mm$^3$. The samples were lapped with diamond paste and etched in hot nitrogen acid, to remove any damage induced by lapping deteriorated surface layers. A standard four-probe *ac* (9Hz) technique was used for resistance measurements. We used Epotek H20E silver epoxy for electrical contacts. Because the sample has a shape of a long rectangular bar its demagnetization factor is nearly zero. A well-defined geometry of the samples provided for the precise $\rho(T)$ and superconducting properties measurements. Temperature was measured with platinum (PT-103) and carbon glass (CGR-1-500) sensors. The measurements were performed in the liquid helium variable temperature cryostat in the temperature range between 1.3 K and 350 K. Magnetic measurements of $\rho(T,H)$ and $\lambda(T,H)$ were carried out using a superconducting coil in applied fields of up to 6 T down to 1.3 K. Additional *dc* and *ac* $\rho(H)$ measurements were performed in the National High Magnetic Field Laboratory in Tallahassee, Flor-ida (NHMFL) at the temperatures down to 0.35 K. The *dc* magnetic field was applied in the direction of the current flow. The critical temperature of the $ZrB_{12}$ samples, measured by RF susceptibility and $\rho(T)$ was found to be $T_{c0}=6.0$ K.

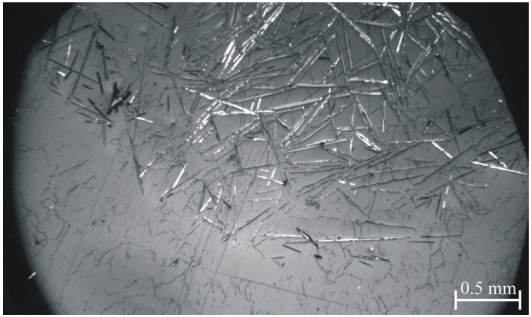

FIG.2. Etching pattern of a $ZrB_{12}$ single crystal cross section in (100) plane from the same parts of ingot as in [19, 23]. The needle like long grains are $ZrB_2$ phase, while the small black lines correspond to etching pits from small angle boundaries.

We used radio frequency LC technique [28] to measure $\lambda(T)$ of $ZrB_{12}$ samples. This technique employs a rectangular solenoid coil into which the sample is placed. The coil is a part of the LC circuit driven by a marginal oscillator operating at 2-10 MHz, or by the impedance meter (VM-508 TESLA, 2-50 MHz). Changes in the properties of the sample lead to the change of the coil's inductance that in turn results in the change of the resonance frequency of the LC circuit. The connection between parameters of the circuit and $\lambda(T)$ is described by following equation:

$$\lambda(T) - \lambda(0) = \delta \cdot \frac{f^{-2}(T) - f^{-2}(0)}{f^{-2}(T_c) - f^{-2}(0)} \quad (1)$$

Here $\delta = 0.5(c^2\rho/2\pi\omega)^{1/2}$ is the imaginary part of a skin depth above $T_c$ [29], which was determined from the $\rho(T)$ measurements close to $T_c$, $f(T)$ is the resonance frequency of the circuit at arbitrary $T$, $f(T_c)$ and $f(0)$ are the resonance frequency of the circuit at the superconducting transition and at zero temperature, respectively.

The $\lambda(T)$ dependence in thin $MgB_2$ films was investigated employing a single coil mutual inductance technique. This technique, originally proposed in [30] and improved in [31], takes advantage of the well known two-coil geometry. It was successfully used for the observation of the Berezinskii – Kosterlitz - Thouless vortex-antivortex unbinding transition in ultrathin $YBa_2Cu_3O_{7-x}$ films [32] as well as for study of the $\lambda(T)$ dependence on $MgB_2$ films [33]. In this radio frequency technique one measures the temperature dependence of the complex impedance of the LC circuit formed with a one-layer pancake coil located in the proximity (~0.1 mm) of the film. Both sample and coil are in a vacuum, but the coil holder is thermally connected with helium bath, while the sample holder is isolated and may be heated. During the experiment the coil was kept at 2.5 K, whereas the sample temperature has been varied from 2.5 K up to 100 K. Such design allows us to eliminate possible effects in temperature changes in L and C on the measurements.

The complex mutual inductance *M* between the coil and the film can be obtained through:

$$\operatorname{Re} M(T) = L_0 \cdot (\frac{f_0^2}{f^2(T)} - 1) \quad (2)$$

$$\operatorname{Im} M(T) = \frac{1}{2\pi \cdot f(T)C^2}[\frac{1}{Z(T)} - \frac{1}{Z_0(T)}\frac{f^2(T)}{f_0^2}] \quad (3)$$

Here L, Z(T), f(T), $L_0$, $Z_0$ and $f_0$ are the inductance, the real part of impedance and the resonant frequency of the circuit with and without the sample, respectively. In the London regime, where the high frequency losses are negligible, one can introduce $\Delta\operatorname{Re}M(T)$ - the difference between temperature dependant real part of *M* of the coil with the sample, ReM(T), and that of the coil at $T_0$, $ReM_o$. This difference is a function of the London penetration depth $\lambda(T)$:

$$\Delta \operatorname{Re} M(T) = \pi\mu_0 \int_0^\infty \frac{M(q)}{1 + 2q\lambda \cdot \coth(\frac{d}{\lambda})} dq \quad (4)$$

where M(q) plays the role of mutual inductance at a given wave number q in the film plane and depends on the sample-coil distance, d is the sample thickness (additional details can be found in [31]). A change in $\Delta\operatorname{Re}M(T)$ is detected as a change of resonant frequency f(T) of the oscillating signal through Eq.2. This change when put into Eq.4 yields temperature dependent London penetration depth $\lambda(T)$. Thanks to Eq.4, we can measure the $\lambda(T)$ of superconducting film by measuring the variation $\Delta\operatorname{Re}M(T)$ of coil impedance and convert them into $\lambda(T)$.

### III. ELECTRON TRANSPORT

Figure 3 shows the temperature dependence of $\rho(T)$ of $ZrB_{12}$ and $ZrB_2$ single crystal samples. To emphasize the variation of $\rho(T)$ in a superconductive state, we plot these data in the inset. The transition temperature ($T_{c0}=6.0$ K) is consistent with the previously reported values for $ZrB_{12}$ (6.03 K) [11, 13] and is larger than that of $ZrB_2$ polycrystalline samples (5.5 K) [2]. The $ZrB_{12}$ samples demonstrate a remarkably narrow transition with $\Delta T=0.04$ K. We believe that such narrow transition is indicator of the good quality of our samples.

Figure 3 does not show any hints of the superconducting transition in $ZrB_2$ single crystals down to 1.3 K [18], even though superconductivity was observed before at 5.5 K in polycrystalline samples [2]. It was recently suggested [33] that this apparent contradiction could be associated with nonstoichiometry in the zirconium sub-lattice. Based on the electron structure calculation it was suggested that the Fermi level in $ZrB_2$ is located in the pseudo gap [34]. The presence of Zr defects in $Zr_{0.75}B_2$ leads to the appearance of a very intense peak in the density of states in the vicinity of the pseudogap and subsequent superconductivity. We strongly believe that observation of superconductivity at 5.5 K in polycrystalline samples of $ZrB_2$ was due to nonstoichiometry of our samples. It is likely that recent observations of superconductivity in nonstoichiometric $Nb_{1-x}B_2$ compounds [4,5] as well as in other nonstoichiometric borides [3,6] have the same origin.

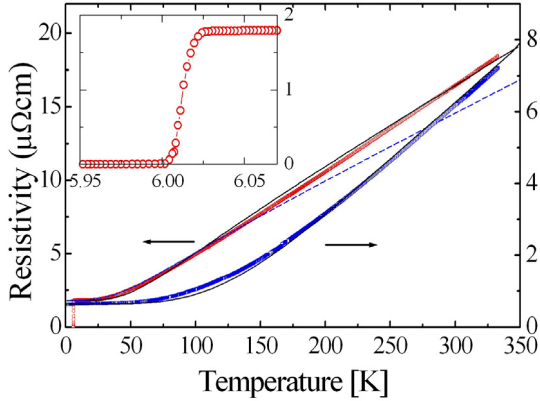

FIG.3. (Color online) Temperature dependence of ρ(T) for $ZrB_{12}$ (circles) and $ZrB_2$ (squares) single crystal samples. The solid lines represent BG fits to the experimental data by Eq.4. The dashed line is $t^3$ BG fit below 170 K with $T_R$=283 K as in [21].

As we can see from Fig.3, despite the fact that $ZrB_{12}$ contains mostly boron, its room temperature ρ(T) is only twice as large as that of single crystal samples of $ZrB_2$. The ρ(T) of $ZrB_{12}$ is linear above 90 K with the slope markedly steeper than in $ZrB_2$, with rather low residual resistivity ratio $\rho_{300K}/\rho_{6K} \approx 10$. One can predict a nearly isotropic ρ(T) dependence for *fcc* $ZrB_{12}$, which can be described by Bloch-Grüneisen (BG) equation of the electron-phonon (*e-p*) scattering rate [35]:

$$\rho(t) - \rho(0) = 4\rho_1 \cdot t^5 \int_0^{1/t} \frac{x^5 e^x dx}{(e^x - 1)^2} = 4\rho_1 \cdot t^5 J_5(1/t) \quad (5)$$

Here, ρ(0) is the residual resistivity, $\rho_1 = d\rho(T)/dt$ is the slope of ρ(T) at high $T>T_R$, $t = T/T_R$, $T_R$ is the resistive Debye temperature and $J_5(1/t)$ is the Debye integral.

It is clear from Fig.3 that the BG model nicely describes the ρ(T) dependence of both borides, indicating the importance of *e-p* interaction. It is remarkable that this description works well with constant $T_R$=300 K, which is very close to $T_R$=280 K observed on polycrystalline samples [16]. Clearly, $ZrB_2$ ($T_R$=700 K) and $ZrB_{12}$ have very different ρ(T) dependence due to different $T_R$. At the same time, the phonon Debye temperature, $T_D$, for $ZrB_{12}$ calculated from C(T) on rather large sample (4.7×4.8×2.9 mm) prepared by one of us (VAG) (without metallographic study) [19,23], is three times higher. Furthermore $T_D$ increases from 800 to 1200 K as temperature rises from $T_c$ up to room temperature. We believe that this inconsistency of $T_R$ and $T_D$ can be explained by limitation of $T_R$ by a cut-off phonon wave vector q = $k_BT/\hbar$. The latter is limited by the Fermi surface (FS) diameter $2k_F$ [36] rather than the highest phonon frequency in the phonon spectrum [18]. Besides some problems may arise in [19, 23] due to use of $ZrB_2$ phase rich samples as in Fig.2 (see below).

Actually, ρ(T) of $ZrB_{12}$ and $ZrB_2$ samples deviates from the BG model at low temperatures [18]. We have been reported for $ZrB_{12}$, $ZrB_2$ and $MgB_2$ that such deviation are consistent with a sum of electron-electron (*e-e*), $aT^2$, and *e-p*, $bT^5$ contributions to the low-T ρ(T) data. The coefficient b= 497.6 $\rho_1/T_R^5$ in this plot, gives another measure of $T_R$ from low-T ρ(T) data. We found this $T_R$ in a good agreement with that extracted from full-T BG fit for both $ZrB_2$ and $ZrB_{12}$ samples. Therefore, the data extracted from this two-term fit are self consistent with the full Eq.5 fit. We would like to stress, that this observation is only possible in the approximation of constant $T_R$, which is in contradiction with specific heat data [19].

Notice also, that the *e-p* contribution to resistivity ρ(T) can be described through [37]:

$$\rho_{ep}(T) \propto \int_0^\infty \alpha^2(k,\omega) F(\omega) \Phi(\varepsilon, \hbar\omega) d\omega, \quad (6)$$

where $\alpha^2(k,\omega)F(\hbar\omega)$ is the effective Eliashberg density of states of the phonons, that is *e-p* coupling function $\alpha^2(k,\omega)$ multiplied by the phonon density of states $F(\hbar\omega)$ with energy $\hbar\omega$ [37]. The $\alpha^2(k,\omega)$ is proportional to the matrix element of *e-p* coupling averaged over phonon polarization, but only longitudinal phonons are responsible for *e-p* scattering for spherical Fermi surface. At the same time, the phonon specific heat can be expressed as:

$$C_{ph} = 3R \int_0^\infty F(\omega) \Phi_1(\hbar\omega) d\omega \quad (7)$$

Here $\Phi(\varepsilon, \hbar)$ and $\Phi_1(\hbar\omega)$ are the occupation factors for *e-p* and phonon systems, respectively [37]. This means, that different phonons are responsible for *e-p* ρ(T) and C(T). In particular, the transverse phonons are much less important for *e-p* scattering whereas both transverse and longitudinal phonons equally contribute to C(T).

Borides have rather high $T_D$ that depresses the *e-p* scattering, as a result *e-e* scattering term may be much more pronounced. Indeed, we find very similar values of *a* − coefficient for $ZrB_{12}$ and $ZrB_2$ samples in the basal plane (a = 22 pΩcmK$^{-2}$ and 15 pΩcmK$^{-2}$, respectively) [18]. It is interesting to note however that these values are five times larger than *e-e* term for transition metals ($a_{Mo}$=2.5 pΩcm/K$^2$ and $a_W$=1.5-4 pΩcm/K$^2$ [18]). In general, there are many scattering processes responsible for the $T^2$ term in ρ(T) of metals. This term could be due to electron-impurity, elec-tron-dislocation scattering, etc. induced deviation from Mattiessen rule. It is difficult to separate the contributions of these effects, thus it is presently not clear where exactly this $T^2$ term comes from [18]. Therefore, additional experiments on more pure samples must be performed, before final conclusion about the origin of the $T^2$ term in borides can be drawn.

As we mentioned in the introduction of this report there is a contradiction between our description of the ρ(T) and that of Refs.19,21. Daghero *et al.* suggested BG fit with $T^3$ dependence rather than $T^5$ on the similar single crystals at low T<170 K [21]. The fit assumed model $\omega^2$ dependence for the $\alpha^2(k,\omega)F(\omega)$ in Eq.6. Such assumption yielded a good fit to their data at T<170 K, with the $T_R$=283 K similar to our value of 300 K. We would like to stress however that there are strong objections to this modified $bT^3$ BG model [18]. The main problem of this approach is that it completely ignores intra sheet small angle e-p scattering responsible for $T^5$ law and takes into account only inter sheet large angle scattering events. No evidence of this model was observed in transition and non-transition metals. To check the approach of Daghero's group we used it for our data. Fig.3 displays the BG fit with ρ(T)-ρ(0) ∝ $t^3 J_3(1/t)$ in Eq.5 at T<170 K (dashed line). It is clear that this fit is far from consistency at higher temperatures. We believe this is indication that ρ(T) cannot be fitted by a modified BG $t^3$ equation in whole temperature range. We suggest that a sum of $T^2$ and $T^5$ contributions to the low-T ρ(T) may be easily confused with a $T^3$ law [18]. Notice that our observation of BG $t^5$ intraband ρ(T) dependence rather then intersheet $t^3$ law, is very important for given below two-$T_c$ model of two-gap superconductivity for ZrB$_{12}$. In fact, this model is right in the limit of zero interband coupling. The two bands coupling will give a single $T_c$.

Lortz *et al.* [19] report temperature dependence of ρ(T) obtained on the samples cut from ones provided by our group. The ρ(T) is nearly identical to our data, however interpretation is different. Lortz *et al.* fitted ρ(T) with the generalized BG formula, using a decomposition into Einstein modes of $\alpha^2(k,\omega)F(\omega)$. Same approach was applied to C(T) data and it was concluded that there are similarities between F(ω) determined from C(T) and $\alpha^2(k,\omega)$ F(ω) from ρ(T). However, the fit to ρ(T) data was obtained using six fitting parameters for $\alpha^2_k F_k$ and ρ(0). We believe such fit is not better than a simple Debye fit (Eq.5) with just two parameters: $T_R$ and ρ(0). Furthermore it is not clear whether Einstein model is applicable to ZrB$_{12}$. Finally, we note that only phonons with a phonon wave vector q=$k_B T/\hbar s$<$2k_F$ can participate in ρ(T) [36]. Thus $T_R$ is limited by the Fermi surface diameter $2k_F$ rather than the highest phonon frequency in the phonon spectrum, which in turn is important for $T_D$. Notice also, that unconventional $T_D(T)$ dependence observed from C(T) data in [19,23] may be due to the sample problems (see Fig.2).

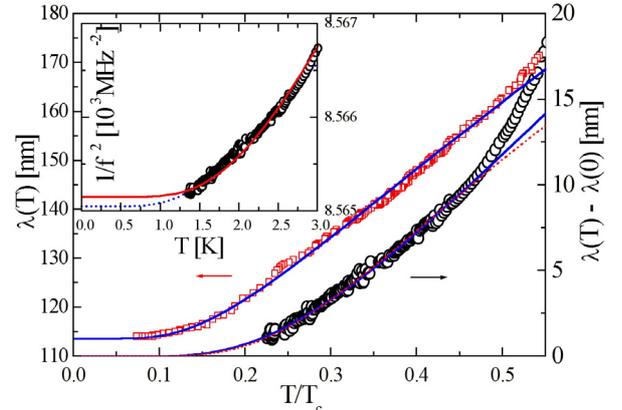

FIG. 4. (Color online) Temperature variation of the λ(T) vs T/T$_c$ for ZrB$_{12}$ single crystal (black circles) and MgB$_2$ thin film (red squares) below 0.55T$_c$. Solid curves represent the single gap dirty limit fit (Eq.9) and dashed line represent the clean limit fit (Eq.8) for ZrB$_{12}$. Inset shows $f^{-2}$(T) and a result of fit by Eq.1 and Eq.9.

## IV. PENETRATION DEPTH

In the BCS theory the London penetration depth is identical with λ(T) for specular and diffuse surface scattering and for negligible nonlocal effects. For a BCS-type superconductor with the conventional *s*-wave pairing form, the λ(T) has an exponentially vanishing temperature dependence below $T_c$/2 (where Δ(T) is almost constant) [38]:

$$\lambda(T) = \lambda(0) \cdot [1 + \sqrt{\frac{\pi \Delta(0)}{2 k_B T}} \cdot \exp(-\frac{\Delta(0)}{k_B T})] \quad (8)$$

for clean limit: l>ξ, and

$$\lambda(T) = \lambda(0) \cdot \sqrt{\frac{1}{\tanh(\frac{\Delta(0)}{2 k_B T})}} \quad (9)$$

for dirty limit: l<ξ [39]. Here Δ(0) is the energy gap and λ(0) is the penetration depth at zero temperature. Close to $T_c$ λ(T) dependence has a BCS form [29]:

$$\lambda(T) = \frac{\lambda(0)}{\sqrt{2 \cdot (1 - \frac{T}{T_c})}} \quad (10)$$

Important problems for λ(T) measurements are: (i) determination of basic superconducting parameter λ(0) and (ii) temperature dependence law, to see whether *s*-wave or *d*-wave pairing form exist. Both these problems can be addressed from low-T λ(T) dependence according to Eq. 8 and Eq. 9. We used Eq.1 to extrapolate the resonance frequency f(T) of our LC circuit down to zero temperature. Inset to Fig.4 shows $f^{-2}$(T) ∝ λ(T) - λ(0) used for determination f(0). We would like to stress, that one can use linear dependence of $f^{-2}(T)$ and hence $\lambda(T) - \lambda(0)$ below 3 K [15-17], due to uncertainty with $f(0)$.

The unconventional *d*-wave pairing symmetry causes the energy gap to be suppressed along nodal directions on the Fermi surface. This type of pairing should manifest itself through the linear temperature dependence of $\lambda(T) - \lambda(0) \propto T$ at low-*T*. Such a linear *T* dependence of $\lambda(T)$ has been used as an indicator of *d*-wave pairing in cuprates [39,40]. At the same time the microwave $\lambda(T)$ data in fully oxygenated YBCO films show a picture which is consistent with the two-band *s*-wave superconductivity [42]. Recently, it was suggested [43] that a strictly linear *T* dependence of $\lambda(T)$ at low temperatures violates the third law of thermodynamics because it results in non vanished entropy in the zero temperature limit. One can argue that a deviation of the linear *T* dependence of $\lambda(T)$ should be observed at low *T*. Indeed, recent experiments on $\lambda(T)$ in cuprates indicate deviations from linearity at low-*T* from current carrying zero energy surface Andreev bound states [44]. We believe that the question about linear dependence of $\lambda(T)$ is still open therefore we use BCS Eq.8 and Eq.9 to fit our data.

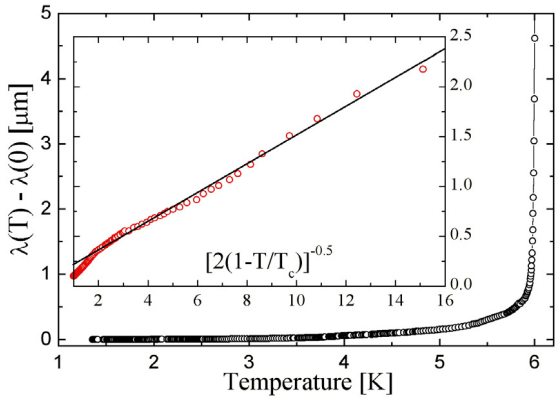

FIG.5. (Color online) Temperature variations of $\lambda(T)$ for $ZrB_{12}$ sample. The inset shows $\lambda(T)-\lambda(0)$ versus BCS reduced temperature.

Dashed curve in inset to Fig.4 is a result of the fit by aid of BCS Eq.9 for $\lambda(T)$ with $f(0)$ and $\Delta(0)$ as free parameters. The Eq.1 defines difference between extrapolated $\lambda(0)$ at zero temperature ant that at the arbitrary temperature *T*, $\lambda(T) - \lambda(0)$. We would like to stress that $\lambda(T) - \lambda(0)$ data are robust relative to the change of the oscillator frequency. We observed no change in data when oscillator frequency has been increased from 2 to 10 MHz. Figure 4 shows how $\lambda(T)-\lambda(0)$ changes with reduced temperature, $T/T_c$, at low-*T* for both $MgB_2$ (black circles) and $ZrB_{12}$ (red circles). Our $ZrB_{12}$ data do not extend to as low reduced temperatures as our data for $MgB_2$. This could lead to somewhat larger uncertainty in the estimates for the zero temperature resonance frequency $f(0)$, and hence $\lambda(0)=66$ nm and $\Delta(0)$ from low-*T* data for $ZrB_{12}$.

To address the problem with $\lambda(0)$ we plot $\lambda(T)-\lambda(0)$ data versus BCS reduced temperature: $1/\sqrt{2(1-T/T_c)}$ in the region close to $T_c$ (see inset to Fig.5). The advantage of this procedure is in the insensitive of such analysis to the choice of $f(0)$ on this temperature scale. The value of $T_c = 5.992$ K used in this data analysis is obtained by getting best linear fit of the $\lambda(T) - \lambda(0)$ versus $1/\sqrt{2(1-T/T_c)}$ plot. Remarkably there is only a few millidegrees difference between $T_c$ obtained from the fit and actual $T_{c0}$. We use the slope of $\lambda(T) - \lambda(0)$ vs $1/\sqrt{2(1-T/T_c)}$ and Eq.10 to obtain the value of $\lambda(0)=143$ nm from. To assure that this $\lambda(0)$ is in agreement with low-T data, we fit the $f^{-2}$ vs T data with Eqs.1 and 9 using fixed $\lambda(0)=143$ nm and free $\Delta(0)$ at low *T*. This fit is shown in inset to Fig.4 by solid curve. It is clear from this inset that high-*T* $\lambda(0)$ is in agreement with low-*T* experimental data.

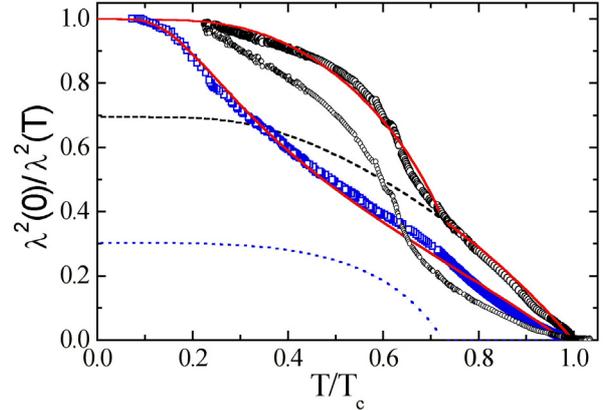

FIG. 6. (Color online) Superfluid density, $[\lambda(0)/\lambda(T)]^2$, of the $ZrB_{12}$ single crystal sample for the $\lambda(0)=143$ nm (open circles) and $\lambda(0)=66$ nm (small circles), and $MgB_2$ thin film (squares). The predicted behavior of $[\lambda(0)/\lambda(T)]^2$ within the two band model as described in the text is shown by the solid, dashed (*p*-band) and dotted (*p*-band) lines. The solid line represent BCS fit of $MgB_2$ data using $2\Delta(0)/k_BT_c$ as fit parameter.

After completer the analysis of the residual penetration depth we attempt to estimate the residual mean free path *l*. In particular we employ Drude formula, ($\rho(0)= 3/N_0 l v_F e^2$) where we use measured $\rho(0)=1.8$ $\mu\Omega$cm, the electron density of states determined from C(T) data, $N_0=1.83\times10^{22}$ st/eV $cm^3$ [23], and the electron Fermi velocity of $v_F=1.9\times10^8$ cm/sec (determined from E(k) data [20]), to obtain $l = 33$ nm. This value is smaller than a coherence length ($\xi(0)=45$ nm, see below) indicating that our sample is in dirty limit. This is confirmed in Fig.4 by slightly better fits of the $\lambda(T) - \lambda(0)$ with Eq.9 up to $T/T_c=0.5$ relative to the clean limit (Eq.8, red curve). In order to investigate the temperature dependence of $\lambda(T)$ in the whole temperature region, in Fig. 6 we plot the superfluid density $\lambda^2(0)/\lambda^2(T)$) versus the reduced $T/T_c$ for $ZrB_{12}$ sample using the $\lambda(0)=143$ nm determined from one gap fit close to $T_c$ (Eq.10), and $\lambda(0)=66$ nm as determined from low T fit (Eq.9), for comparison.

One can easily notice from Fig.6 an unconventional behavior of $ZrB_{12}$ superfluid density with pronounced shoulder at $T/T_c$ equal to 0.65 for both $\lambda(0)$. This feature can be explained by a model of two independent BCS superconducting bands with different plasma frequencies, gaps and $T_c$'s [24]. We label these two bands as $p$- and $d$-bands according to electron structure of $ZrB_{12}$ [20]. Assuming parallel currents through alternating subsystems, the conductivity is a sum of partial bands conductivities. The imaginary part of the conductivity is proportional to $1/\lambda^2$. For a dirty limit [45] we can write in:

$$\frac{1}{\lambda^2(T)} = \frac{\Delta_p(T)\cdot\tanh(\frac{\Delta_p}{2k_BT})}{\lambda^2_p(0)\Delta_p(0)} + \frac{\Delta_d(T)\cdot\tanh(\frac{\Delta_d}{2k_BT})}{\lambda^2_d(0)\Delta_d(0)} \quad (11)$$

Here $\Delta_i$ is the superconducting energy gap and $\lambda_i(0)$ is residual penetration depth in $p$- or $d$-band. Using this two gap $\lambda(T)$ BCS-like dependence and interpolation formula $\Delta(T)=\Delta(0)\tanh[1.88(T_c/T-1)^{1/2}]$ we fit the experimental data with six fitting parameters: $\lambda_i$, $\Delta_i$ and $T_{ci}$. From this fit we obtain $T^p_c= 6.0$ K, $T^d_c=4.35$ K, $\Delta_p(0)=0.73$ meV, $\Delta_d(0)=1.21$ meV, $\lambda_p(0) =170$ nm and $\lambda_d(0) =260$ nm, for $p$ and $d$ bands, respectively. Dashed and dotted lines in Fig. 6, show the contributions of each $p$- and $d$- bands, respectively. Notice, that this analysis was applied for more reliable $\lambda(0)=143$ nm data. Clearly low temperature dependence of $\lambda^2(0)/\lambda^2(T)$ is dominated by the $d$-band with the smallest $T_c$, whereas the high temperature behavior results from the $p$- band with the larger $T_c$. The reduced energy gap for $p$- band, $2\Delta_p(0)/k_BT_c^p = 2.81$, is rather small relative to the BCS value 3.52, while $d$- band value, $2\Delta_d(0)/k_BT_c^d =6.44$, is twice as big. Thus, we suggest that $ZrB_{12}$ may have two superconducting bands with different $T_c$ and order parameters. Notice, that this unusual conclusion may be right for two bands in the limit of zero interband coupling in agreement with resistivity data. Also, the $\Delta(0)$ of $ZrB_{12}$ may not be constant over the Fermi surface. Here we just tried to fit a data assuming a two gap distinct values.

We based our conclusion on the two-gap model for dirty-limit superconductors, suggested by Gurevich [45]. In this model, we can write:

$$\frac{1}{\lambda^2(T)} = \frac{4\pi^2 e^2}{\hbar c^2}(N_p\Delta_p D_p + N_d\Delta_d D_d) = \quad (12)$$
$$= \frac{1}{\lambda_p^2(0)} + \frac{1}{\lambda_d^2(0)}$$

where $N_i$, $\Delta_i$ and $D_i$ are the density of states, the energy gap and the diffusivity in $p$- and $d$- bands, respectively. The calculated band structure of $ZrB_{12}$ is composed of three 3D Fermi surface sheets: an open sheet along $\Gamma L$ direction with $k_{\Gamma X}=0.47$ Å$^{-1}$, a quasi spherical sheet at point X ($k_{X\Gamma} = 0.37$ Å$^{-1}$) and a small sheet at point K ($k_{K\Gamma} = 0.14$ Å$^{-1}$) [20]. It is not clear from [20] whether the wave functions of carriers on these sheets are due to predominantly $p$- or $d$- states. However it follows, that the dominant contribution to the density of states $N(E_F)$ is made by the $Zr_{4d}$ and $B_{2p}$ states, with $N_d = 7.3\times10^{21}$ st./eVcm$^3$ and $N_p = 8.7\times10^{21}$ st./eV cm$^3$, respectively [20]. The $B_{2p}$ bonding states are responsible for the formation of $B_{12}$ intra-cluster covalent bonds. In turn, $Zr_{4d}$ bands are due to Zr sub-lattice. A much smaller boron isotope effect on $T_c$ in comparison with Zr isotopic substitution [13,14] may be an indication of the existence of two separate subsystems with different gaps and $T_c$ values. We use this two band approach (Eq.11) to obtain $p$-band diffusivity of $D_p$=57 cm$^2$/sec and $d$-band diffusivity of $D_d$=10 cm$^2$/sec. Note that there is almost a six times difference between $p$-and $d$-band diffusivity. We use this result for our discussion of $H_{c2}(T)$ data in the following paragraph.

The important goal of this paper is comparison of $ZrB_{12}$ and $MgB_2$ data. In Fig.4 we show the temperature variation of $\lambda(T)$ at $T<T_c/2$ and in Fig.6 a superfluid density, $\lambda^2(0)/\lambda^2(T)$, versus reduced temperature $T/T_c$ for the best $MgB_2$ film as determined from the one-coil technique (Eq.2) and inversion procedure from Eq.4 with $\lambda(0)=114$ nm. The solid line represent BCS single gap calculations by aid of single term of Eq.11 and using finite energy gap $\Delta(0)=1.93$ meV as fit parameter. According to Fig. 6, there is a very good agreement between experimental data and the BCS curve over the full temperature range. Simple conventional $s$-wave dirty case fit by Eq.9, agrees remarkably well with the low-T data at $T<T_c/2$ too. The reduced energy gap $2\Delta(0)/k_BT_c$ is evaluated to be 1.14. It is actually within the range of values for 3D $\pi$ - bands obtained by PCS on $MgB_2$ single crystals ($\Delta_\pi$= 2.9 meV) [46], and it agrees with $\Delta(0)$ data obtained from similar radio-frequency experiments on single crystals (1.42 meV [47]), thin films (2.3 meV [32]) and with the theoretical prediction of the two-band model [44]. Rather small $2\Delta(0)/k_BT_c$ observed correspond to the small energy gap in the two-gap model for 3D $\pi$ - band. Notice, that we studied here the penetration depth in the $ab$ plane due to the samples being c-axis oriented thin films. This feature predicts that our $\lambda_{ab}(T)$ is determined by the small energy gap for $\pi$ - band. Both $\Delta(0)$ and $\lambda(0)$ are consistent with microwave measurements on similar c-axis oriented thin films (3.2 meV and 107 nm, respectively) [48].

## V. UPPER AND LOWER CRITICAL MAGNETIC FIELD

We now turn to the electronic transport data acquired in magnetic field. We measured the depend-ence of ρ on magnetic field $H$ in the temperature range between 0.35 K and 6 K in two different magnets at NHMFL, Tallahassee, FL as well in our superconducting coil. First magnet was resistive coil Bitter magnet, and second one was a superconducting magnet. Fig.7 displays the resistive magnetic field transitions at various temperatures down to 0.35 K with the fields oriented along the sample bar. Two features are clearly seen: (i) as temperature decreases the resistive transition continuously moves to higher fields without any saturation, (ii) longitudinal magnetoresistivity in the normal state is very small. We used several different approaches to extract $H_{c2}(T)$ from our data. Initially, we extended the maximal derivative $d\rho/dH$ line (dashed line in Fig.7) up to the normal state resistivity $\rho(T)$ level. The crossing point of this line and the normal state resistivity gave us the estimate for $H^*_{c2}$ at various temperatures, as is indicated by the arrow. Despite a clear broadening at the higher fields, the onset of the resistive transition remains well defined even at rather low temperatures. To obtain an alternative estimate for $H_{c2}$, we fitted the field dependence of $\rho(H)$ close $H_{c2}$ by a cubic polynomial and calculated the derivative $d\rho/dH$. We defined $H_{c2}$ as the field where $d\rho/dH$ just starts to deviate from zero (see Fig. 7).

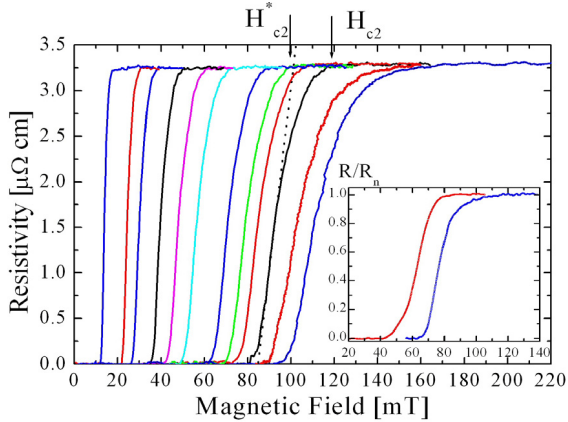

FIG.7. (Color online) Resistivity of $ZrB_{12}$ in the vicinity of the superconducting transition as a function of $H$ at different $T$: 5.45 K, 5.08 K, 4.9 K, 4.26 K, 3.72 K, 3.26 K, 2.59 K, 2.16 K, 1.85 K, 1.39 K, 1.05, 0.35 K from the left to the right. The dotted line and the arrows describe how the $H_{c2}$ has been established. The inset shows the $R(H)/R_n$ near the transition at T=2.3 K for the same sample (blue line) and a sample cut from $ZrB_2$ rich part of an ingot (red line).

It is important to mention that even in single crystalline samples the resistance can be affected by the defects and surface superconductivity. To get even better fill for $H_{c2}$ we used our $\lambda(H)$ data. Figure 8 shows a plot of the $\lambda$ versus the longitudinal magnetic field $H$ measured at various temperatures. (We use the same rectangular coil LC technique as for zero field $\lambda(T)$ measurements.) To avoid demagnetization effects as in [21,22], we oriented our bar-shape sample with its longer side parallel to the external DC field. Changes in the magnetic field dependence of $\lambda(H)$ are directly proportional to the RF susceptibility of the sample and reflect the bulk properties of it. To deduce the $H_{c2}$ from $\lambda(T)$, we used approaches similar to those applied to $\rho(T)$ data, i.e. a straight-line fit representing the maximum of derivative $d\lambda/dH$ (dashed line in Fig.8) was extended up to the normal state. $H^*_{c2}$ was defined as a crossing point of this line with normal state skin depth δ. Fig.8 clearly demonstrates a well defined onset of $\lambda(H)$ transition. We used this onset to estimate $H_{c2}$.

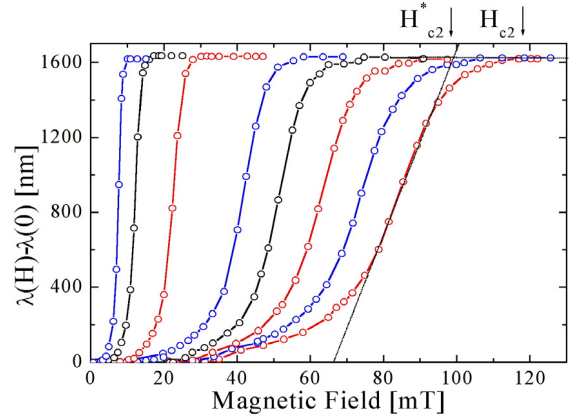

FIG. 8. (Color online) Magnetic field variation of $\lambda(H)$ of a single crystal $ZrB_{12}$ sample at different temperatures: 5.66 K, 5.53 K, 5.02 K, 4.06 K, 3.45 K, 2.84 K, 2.15 K and 1.43K, from the left to the right. The solid lines are the guides for the eye. The dotted line and the arrows describe how $H^*_{c2}$ and $H_{c2}$ has been deduced.

Figure 9 shows the magnetic field $\lambda(H)$ behavior at very small fields. The $\lambda(H)-\lambda(0)$ curves display clear linear dependence at low fields caused by the Meissner effect. We determined the value of $H_{c1}$ from crossing point of two linear dependences below and above the break point on $\lambda(T)$ (see Fig.9).

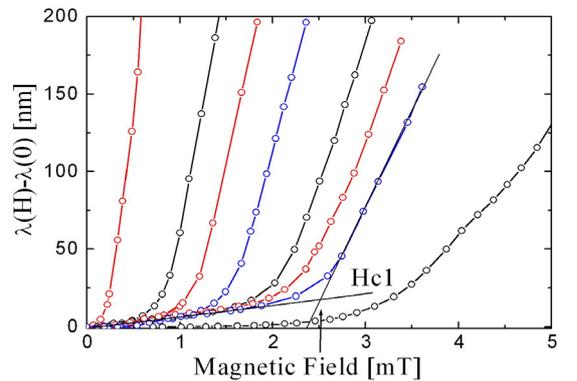

FIG. 9. (Color online) Low magnetic field variation of $\lambda(T)$ of a $ZrB_{12}$ sample at small fields at various temperatures: 5.8 K, 5.2 K, 4.8 K, 4.4 K, 3.8 K, 3.5 K, 3.1 K, 1.5 K, from the left to the right. The solid lines are the guides for the eye. The dotted line is the linear extrapolation of the data used for $H_{c1}$ determination.

Figure 10 presents the $H^*_{c2}(T)$ dependence obtained from extrapolation of the maximum of slopes of both $\rho(H)$ and $\lambda(H)$, as well as those defined at the onsets of the finite $\rho(H)$ and $\lambda(H)$. In the same figure we also plot the $H_{c1}(T)$, acquired by using the break point of $\lambda(H)$ criteria as the definition of the lower critical magnetic field. Remarkable feature of this plot is an identical linear increase of $H_{c2}$ with decreasing temperature for each of the methods of defining $H_{c2}$. As we can see from this figure, the $H_{c2}$ data obtained at three different magnets agree remarkably well and are aligned along corresponding straight lines indicating linear $H_{c2}(T)$ dependence down to 0.35 K.

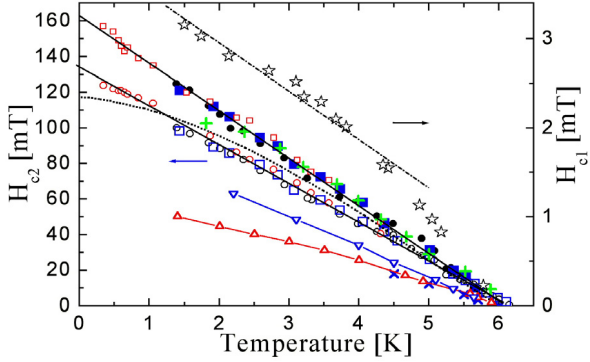

FIG.10. (Color online) Temperature variations of $H_{c2}(T)$ and $H_{c1}(T)$ (stars) of $ZrB_{12}$. Symbols: $H^*_{c2}(T)$ determined from $\rho(H)$ (circles) and $\lambda(H)$ (squares) extrapolations; closed points – the onset $H_{c2}(T)$ data, as described in the text. Open red circles and squares are $\rho(T)$ data obtained in NHMFL. Dotted line is the BCS $H_{c2}(T)$ data determined from the onset data and HW formula [50]. The straight crosses - PCS data [21], tilted crosses - magnetization data [22], up red triangles - C(T) data, and down blue triangles – the $\rho(H)$ data of [19, 23].

To see whether one gap BCS model may work for $ZrB_{12}$, we extrapolate $H_{c2}(T)$ to zero temperature by use of the derivative of $dH_{c2}(T)/dT$ close to $T_c$ and the assumption that the zero temperature $H_{c2}(0) = 0.69T_c dH_{c2}/dT|_{Tc}$ [38]. The resulting $H_{c2}(0)$=114 mT is substantially lower than the low temperature onset data below 3 K (see Fig.10). Linear extrapolation of $H_{c2}(T)$ to T=0 gives $H_{c2}(0) = 162$ mT. This value is almost the same that obtained in the polycrystalline $ZrB_{12}$ samples (150 mT) [16]. We used this value to obtain the coherence length $\xi(0)$, by employing the relations $H_{c2}(0) = \phi_0/2\pi\xi^2(0)$. The latter yields $\xi(0) = 45$ nm, which is substantially larger than a few angstroms coherence length of high-$T_c$ superconductors.

In contrast to Ref. 19, 22 and 23 our estimations agree well with the Ginzburg-Landau parameter $\kappa=\lambda/\xi$. Using our values of $\lambda_p$ and $\lambda_d$ data we obtain $\kappa_p$=3.8 and $\kappa_d$=5.8. Both values of $\kappa_d$ are larger then $1/\sqrt{2}$ that implies that $ZrB_{12}$ is type II superconductor at all $T$. Using the GL expression for $H_{c1}(T) = \phi_0 \ln\kappa /4\pi\lambda^2$ we obtain: $H_{c2}/H_{c1}= 2\kappa^2/\ln\kappa$. From the value $H_{c2}(0)$ obtained above and the $H_{c1}$ data from Fig. 9, we find $\kappa$=6.3 and $\lambda(0)$=280 nm which is in good agreement with the value $\lambda_p(0)$=260 nm obtained from two gap BCS fit for $d$- band. One could argue that the $H_{c1}(T)$ obtained from our magnetic field measurements of $\lambda(T)$ may reflect the flux entry field because of the Bean-Livingston surface barrier, rather then trough $H_{c1}$. However the entry field $H_{BL} = H_{c1}\kappa/\ln\kappa$ [49] is three times larger than $H_{c1}$ even when one uses $\kappa$=6.3, so if we assume $\kappa$=23 given by a ratio of $H_{c2}/H_{c1}$ we obtain zero temperature $\lambda(0)$=1030 nm, which is unreasonably large compared to one obtained from a BCS fit. This result is another confirmation of our suggestions.

In contrast to the conventional BCS theory [50], $H_{c2}(T)$ dependence is linear over an extended temperature range with no evidence of saturation down to 0.35 K. Similar linear $H_{c2}(T)$ dependence have been observed in $MgB_2$ [51,52] and $BaNbO_x$ [53] compounds. One can describe this behavior of upper critical filed using two-gap approach. According to Gurevich [45], the zero-temperature value of the $H_{c2}(0)$ is significantly enhanced in the two gap dirty limit superconductor model:

$$H_{c2}(0) = \frac{\phi_0 k_B T_c}{1.12\hbar\sqrt{D_1 D_2}} \exp(\frac{g}{2}) \quad , \qquad (12)$$

as compared to the one-gap dirty limit approximation $H_{c2}(0)=\phi_0 k_B T_c/1.12\hbar D$. Here $g$ is rather complicated function of the matrix of the BCS superconducting coupling constants.

In the limit of $D_2<<D_1$ we can simply approximate $g\approx|\ln(D_2/D_1)|$. Large ratio of $D_2/D_1$ leads to the enhancement of $H_{c2}(0)$ and results in the upward curvature of the $H_{c2}(T)$ close to T=0 [45]. According to our $\lambda(T)$ data (see above), we found very different diffusivities for $p$- and $d$- bands: $D_p/D_d \approx 3$. Thus we can speculate that the limiting value of $H_{c2}(0)$ is dominated by $d$- band with lower diffusivity $D_d$=17 cm$^2$/sec, while the derivative $dH_{c2}/dT$ close to $T_c$ is due to larger diffusivity band ($D_p$=56 cm$^2$/sec). Indeed, simple estimation of $D_p$=4$\phi_0 k_B/\pi^2\hbar dH_{c2}/dT$ =39 cm$^2$/sec from derivative $dH_{c2}/dT$=0.027 T/K close to $T_c$, gives almost the same diffusivity relative to one estimated from $\lambda(T)$ for p-band. Thus, we believe that two gap theoretical model of Gurevich, qualitatively explains the unconventional linear $H_{c2}(T)$ dependence, which supports our conclusion about two gap nature of superconductivity in $ZrB_{12}$.

The possibility of the multigap nature of the superconducting state was predicted for a multiband superconductor with large difference of the $e$-$p$ interaction at different Fermi-surface sheets (see [44] and references therein). To the date $MgB_2$ has been the only compound with the behavior consistent with the idea of two distinct gaps with the same $T_c$. We believe that our data can add $ZrB_{12}$ as another unconventional example of multi gap and multi-$T_c$ superconductor. This conclusion contradicts several exist-ing publications however we believe we can successfully defend our idea.

The $H_{c2}(T)$ dependence of $ZrB_{12}$ single crystals from the same Kiev group has been measured by three different groups mentioned above and are plotted in Fig.10. Clearly, our $H_{c2}(T)$ data are very similar to those obtained by Daghero *et al.* [21] (straight crosses). The agreement is nearly perfect except the last data point at 1.8K. The disagreement is in the interpretation. The authors of Ref. 21 concluded that $ZrB_{12}$ is a conventional one gap s-wave superconductor with $\Delta(0)=1.22$ meV. Thus a strong coupling scenario with reduced energy gap of $2\Delta(0)/k_BT_c=4.8$ was proposed. One should note however, that the gap signature in PCS data has been observed in a temperature range close to $T_c$ (4.2-6K), although the second gap signature feature should have been seen only below $T_c^d=4.35$ K, and could have been simply missed by the authors of Ref. 21 because of the limited temperature range of their measurements.

There is nearly two times difference between $H_{c2}$ obtained from our data and that obtained from tunneling and magnetic characterization data of Tsindlekht *et al.* [22]. Same can be said about $H_{c2}$ obtained from specific heat and resistivity data of Lortz *et al.* [19] and Wang *et al.* [23]. Tsindlekht *et al.* concluded that $ZrB_{12}$ is a type-II superconductor with the Ginzburg-Landau parameter κ slightly above the marginal value of $1/\sqrt{2}$. At the same time Wang *et al.* [23] observed a crossover between type I and type II behavior of κ(T) at 4.7 K in contrast to our data. In all those approximations, the validity of the one gap BCS picture is implicitly assumed. Large contradiction between the ρ(H) and C(T) data (see Fig. 10) was attributed to surface superconductivity. One can mention another possibly reason for this contradiction. In particular, data of [23] have been measured on the samples cut by diamond saw without chemical etching of surface damaged layer. This procedure may create a strong concentration gradient of boron in a surface layer and a substantial manifestation of the surface barriers for the flux lines resulting in this contradiction. One can also note that the discrepancies in $H_{c2}$ data can be due to potentially large non homogeneity of the $ZrB_{12}$ samples. Indeed, Fig.7 inset clearly demonstrates different transitions for our sample with no inclusions of $ZrB_2$ and the sample cut from $ZrB_2$ reach part of an ingot (red line).

We believe that this inconsistency of our data and the data of Refs. 19,21-23 can be: (i) due to the two gap nature of superconductivity in $ZrB_{12}$, (ii) due to large uncertainty in determining of the zero temperature gap from very narrow temperature range of the measurements and (iii) last but not least due to potentially large non homogeneity of the $ZrB_{12}$ samples that have been used by other authors (see Fig.2). Although observed two gap behavior of $\lambda^2(0)/\lambda^2(T)$ in $ZrB_{12}$ is similar to that in high-$T_c$ superconductors, observation of two different $T_c$ in these bands is unconventional. This also relates to the linear $H_{c2}(T)$ dependence in the wide temperature range up to $T_c$. Striking two gap BCS behavior observed calls certainly for a new study of low-$T$ energy gap and $H_c(T)$ of $ZrB_{12}$ for understand the nature of superconductivity in this cluster compounds.

## VI. CONCLUSION

We performed systematic study of the temperature and magnetic field dependencies of the resistivity, penetration depth, lower, $H_{c1}(T)$, and upper, $H_{c2}(T)$, critical magnetic fields of the single crystals dodecaboride $ZrB_{12}$ and resistivity of diboride $ZrB_2$, as well as the penetration depth in thin films of $MgB_2$. While the temperature dependence of λ(T) in thin c-axis oriented thin film $MgB_2$ samples is well described by an isotropic *s*-type order parameter, we find unconventional behavior of $ZrB_{12}$ superfluid density with pronounced shoulder at $T/T_c$ equal to 0.65. The $H_{c2}(T)$ dependences have been deduced from the ρ(H) and λ(H) data. Both techniques reveal an unconventional linear temperature dependence of $H_{c2}(T)$, with a considerably low value of $H_{c2}(0) = 0.16$ T. We conclude therefore that $ZrB_{12}$ presents an evidence of the unconventional two-gap superconductivity with different $T_c$ in the different bands.

## VII. ACKNOWLEDGEMENT

We would like to thank V.F. Gantmakher, A. Gurevich, R. Huguenin, D. van der Marel, I.R. Shein, for very useful discussions, V.B. Filipov, A.B. Lyashenko and Yu.B. Paderno for preparation of $ZrB_2$ and $ZrB_{12}$ single crystals, Hong-Ying Zhai, H.M. Christen, M.P. Paranthaman and D.H. Lowndes for providing us of high quality $MgB_2$ films, A. Suslov for help in low-*T* measurements in NHMFL, L.S. Uspenskaja for help in metallographic analysis and L.V. Gasparov for help in paper preparation. This work was supported by the Russian Council on High-Temperature Superconductivity (Grant Volna 4G), the Russian Scientific Programs: Surface Atomic Structures (Grant No.4.10.99) and Synthesis of Fullerens and Other Atomic Clusters (Grant No. 541-028), Russian Ministry of Industry, Science and Technology (Grant MSh-2169.2003.2), the Russian Foundation for Basic Research (Grant No. 02-02-16874-a) and by the INTAS (Grant No. 2001-0617). A portion of this work was performed at the National High Magnetic Field Laboratory, which is supported by NSF Cooperative Agreement No.DMR-0084173 and State of Florida.